\documentclass[12pt]{article}

\newif\ifnew
\newtrue

\usepackage{cite}
\usepackage[pdftex]{graphicx}
\usepackage{amsmath}
\usepackage{algorithmic}
\usepackage{array}
\usepackage{caption} 
\usepackage{amsmath}
\begin{document}

\title{      Unconditionally secure ciphers with a short key for a source with unknown statistics}
\author{ Boris Ryabko  \\ \\{\small Federal Research Center for Information and Computational Technologies } \\ {\small Novosibirsk State  University} \\ }
\date{}
\maketitle

\thispagestyle{empty}

\begin{abstract}

We consider the problem of constructing an unconditionally secure cipher with a short key for the case where the probability distribution of encrypted messages is unknown.  Note that unconditional security means that an adversary with no computational constraints can obtain only a negligible amount of information ("leakage") about an encrypted message (without knowing the key).

Here we consider the case of a priori (partially) unknown message source statistics. 
More specifically, 
 the message source probability distribution belongs to a given family of distributions.
  We propose an unconditionally secure cipher for this case.   As an example, one can consider constructing 
  a  single cipher for texts written in any of the languages of the European Union.   That is, the message to be encrypted could be written in any of these languages.

\end{abstract}

\textbf{Keywords:} { \it  
cryptography,    unconditionally secure cipher,   entropically-secure symmetric encryption scheme,  indistinguishability,  data compression, 
  universal code.}

\section{Introduction }

The concept of unconditional security is very attractive to cryptography and has found many applications since C.~Shannon described it in his famous   article \cite {sh}.  The concept refers to secret-key cryptography involving three participants Alice, Bob and Eve, where Alice wants to send a message to Bob in secret from Eve, who has the ability to read all correspondence between Alice and Bob. To do this, Alice and Bob use a cipher with a secret key $k$ (i.e. a word from some alphabet), which is known to them in advance (but not to Eve).  When Alice wants to send some message $m$, she first encrypts $m$ using key $k$ and sends it to Bob, who in turn decrypts the received encrypted message using the key $k$.   Eve also receives the encrypted message and tries to decrypt it without knowing the key. The system is called unconditionally secure, or perfect, if Eve, with computers and other equipment of unlimited power and unlimited time, cannot obtain any information about the encrypted message.   
Not only did C.~Shannon provide a formal definition of perfect (or unconditional) secrecy, but he also showed that the so-called one-time pad (or Vernam cipher) is such a system. One of the specific properties of this system is the equivalence of the length of the secret key and the message (or its entropy). Moreover, C.~Shannon proved that this property must be true for any perfect system. Quite often this property has limited practical application as many modern telecommunication systems forward and store megabytes of information and the requirement to have secret keys of the same length seems to be quite stringent.  
There are, therefore, many different approaches to overcoming this obstacle. These include the ideal systems proposed by C.~Shannon \cite{sh}, the so-called honeycomb cipher proposed by Jewels and Ristenpart \cite{ho1}, the so-called entropy security proposed by Russell and Wang \cite{fool} and some others developed in recent decades  \cite{ho2,do,med,ca,ry1,ry-vernam,fo2,rya-new}.   

The present work is concerned with entropically secure ciphers. 

It is important to note that an entropically secure cipher is not perfect, and Eve may obtain some information  about the message~--- the property referred to as ``leakage,''   see the definition below, but this leakage can be made negligible. 
On the other hand,  an entropically secure cipher makes it possible  to significantly reduce the key length (compared to the perfect cipher).

Recently, an entropically secure cipher has been proposed for the case where encrypted messages have a known distribution,
and for the case where  messages are generated by a Markov chain \cite{rya-new}.     
In the case of a known distribution, the length of the secret key is independent of message length, while in the case of a Markov chain, the length of the key grows logarithmically with message length;  in both cases the length of the key depends on the amount of leakage. 

In this paper we consider the situation where encrypted messages obey an unknown (or partially unknown) probability distribution.   
We propose an entropically secure cipher for which the key length depends on universal code (or data compressor) used for encoding the source 
and  on the admissible leakage of the cipher. 
In a sense, the problem under consideration includes as special cases the previously solved problems with known probability distribution and the case where messages are generated by a Markov chain.

The construction of the cipher is based on    entropically secure ciphers   \cite{fool,do,fo2,rya-new} and universal coding   \cite{kr}.
It is worth noting that the proposed cipher uses data compression and randomisation, both of which are quite popular in unconditional security, cf. \cite{p1,p2,rf} and \cite{ gu,rf},  respectively.

\section{Definitions and preliminaries}

\subsection{Basic concepts}

We consider the problem of symmetric encryption, where   Alice wants  to securely transmit a  message   to Bob.
The messages are $n$-letter binary words,   they  obey a certain probability distribution $p$ defined on the set $\{0,1\}^n, n \ge 1$.  This distribution is only partially known,  i.e. it is known that $p$ belongs to some given 
set $P$, $P \subset R^n$.  
 Alice and Bob   have a shared secret key $ K = K_1 ... K_k $, and 
Alice encrypts the message $M \in \{0,1\}^n$ using $K$ and possibly some random bits.  Then she sends the word $cipher(M,K)$   to Bob, who decrypts the received $cipher(M,K)$ and obtains $M$.  The third participant is a computationally unconstrained adversary Eve, who knows $cipher(M,K)$ and distribution $p$, and wants to find some information about $M$ without knowing $K$.

Russell and Wang  \cite {fool} suggested  a definition of entropic security which was generalised by Dodis and Smith \cite{do} as follows: 
A probabilistic map $Y$ is said to hide
all functions on $\{0,1\}^n$   with leakage $\epsilon$ if, for every adversary $A$, there exists some adversary $\hat{A}$ (who does not know $ Y(M)$) 
such that for all functions $f$, 
\begin{equation}\label{entrsec} 
| \, Pr\{A( Y(M) ) \, = f(M) \}  \, - Pr\{  \hat{A}(\,)\, = f(M) \} \, | \,\le\, \epsilon .
 \end{equation}
 (note that $\hat{A}$ does not know $ Y(M)$  and, in fact, she guesses the meaning of the function $f(M)$.) 
 In what follows, the probabilistic map $Y$ will be $cipher(M,K)$
 and $f$ is a map
 $f: \{0,1\}^n \to \{0,1\}^*$.
 
{\bf Definition 1.}   
The map $Y ()$ is called $\epsilon$-entropically secure for family probability distributions $P$ if $Y ()$ hides all functions on $\{0,1\}^n$ with leakage of $\epsilon$, whenever $p \in P$.

Note that, in a  sense,  Definition 1  is a generalisation  of  Shannon's notion 
of  perfect security.  Namely, if we take $\epsilon = 0$ and $Y$  $= cipher(M,K)$ and $f(x) = x$, 
we obtain that for any $M$ 
$$ | Pr\{{A}( cipher (M,K) ) = M \}  - Pr \{ \hat{A} (\,\,)  = M \} \,| \, = \, 0 $$
So, $A$ and $\hat{A}$ obtained the same result, but $A$
estimates  the probability based on $cipher (M,K) $, whereas $\hat{A}$ does it without knowledge of  $cipher (M,K) $.
Thus, the entropic security (\ref{entrsec}) can be considered as a generalisation of the Shannon's perfect secrecy. 
  
We will use another important concept, the notion of indistinguishability.   
 
 {\bf Definition 2}     
A randomised map $Y: \{0, 1\}^n \to \{0, 1\}^n, n \ge 1,   $ is $\epsilon$-indistinguishable for some family of destributions $\bf{P}$  and $\epsilon > 0$ if there is a probability distribution  $ G$ on $\{0, 1\}^n$ 
such that for every probability  
distribution  $p \in \bf{P}$   
we have
$$SD(Y (M), G)  \le \epsilon,$$
where for two distributions $A,B$  
$$ SD(A,B) = \frac{1}{2} \sum_{U \in \bf{M}}  | Pr\{A = U\} - Pr\{B = U \} | \, .
$$
Importantly, $G$ is independent of $Y(M)$.
  
 Dodis and Smith \cite{do} showed that the concepts of $\epsilon$-entropic security and $\epsilon$-indistinguishability  are equivalent up to small parameter changes.

\subsection{$\epsilon$-entropically  secure ciphers for distributions with bounded min-entropy}

In 2006  \cite{fool}, the first entropy secure cipher was  developed for probability distributions with a limited value of the so-called minimum entropy,  which is defined as follows
  \begin{equation}\label{entr}
h_{min}(p) =  - \log \, \, \max_{a \in A} \, p(a) \, .
 \end{equation}
where $p$ is a probability distribution, $\log = \log_2 \,$.
The Russell and Wang \cite{fool} cipher was generalized and developed by Dodis and Smith  \cite{do} and their result can be formulated as follows:

  {\bf Theorem  1 \cite{do}}. 
{\it Let $p$ be a probability distribution on  $ \{0,1\}^n, n >0,$ 
 whose min-entropy is not less then $h, h \in [0, n]. $
 Then there exists an    $\epsilon$-entropically secure
cipher with the $k$-bit key where 
\begin{equation}\label{maineqv}
k = n - h + 2  log (1/\epsilon) +2.
\end{equation} }
Let's denote this cipher as $cipher_{rw-ds}$.

In a sense, this cipher generalizes the perfect Shannon cipher as follows: In a perfect cipher the key is the word from $\{0,1\}^n$, while in an entropically secure cipher the key belongs to the $2^k$-element subset $K \subset$ $\{0,1\}^n$, which is a so-called small-biased set.  Informally, this means that for any $m \le n$ and a uniformly chosen binary word $u \in \{0,1\}^m$, for any $m$ positions $i_1 i_2, . . . , i_m$, the probability that $K_{i_1}, K_{i_2}, . . . . , K_{i_m} = u$ is close to $2^{-m}$. (This construction is based on some deep   results 
  in combinatorics \cite{1,2,do}.) Thus, the key length decreases from $n$ to $k$.  
Note that the leakage $\epsilon$ and hence the summand $2  \log (1/\epsilon) +2$ depends on the size of the ``small-biased set''  $2^k$ (In general, larger  $k$ implies smaller $\epsilon$.) 

\subsection{$\epsilon$-entropically secure ciphers with reduced secret key}

In equality (\ref{maineqv}),    the linearly increasing summand $n - h $ depends on the min-entropy $h$. 
 So, it seems natural to transform the set $\{0,1\}^n$ so as to reduce the min-entropy of the original distribution $p$ and hence the summand $n-h$.  In \cite{rya-new} this 
  approach  was realised as follows: let there be a set of probability distributios ${\bf P} $ defined on $\{0,1\}^n, n \ge 1$.  The key part of the  cipher is such a randomised  map 
 $\phi :   \{0,1\}^n   \rightarrow   \{0,1\}^{n^*}, n^* \ge n$,  that  there exists a map $\phi^{-1}$  (i.e $ \forall \, 	u \,\,$ $\phi^{-1}(\phi(u) )= u$)  and a min-entropy of the transform probabiity distribution $\pi_p$ is close to $n^*$
(here the distribution $\pi_p$ is such that  $p(u)  = \sum_{v:    \phi^{-1} (v) = u}     \pi_p(v) $). And then the $cipher_{rw-ds}$ can be applied to $\phi(m)$ with a shorter key, because the difference $n^* - h_{min}(\pi_p)$ will be less that $n - h_{min}(p)$,
see (\ref{maineqv}).  
Thus, the smaller $\sup_{p \in P} (n^* - h_{min}(\pi_p) )$, the shorter the secret key. The described cipher is based on data compression and randomisation and denoted in \cite{rya-new} by $cipher_{c\&r}$.   The following theorem describes its properties.

 {\bf Theorem  2 \cite{rya-new}}. {\it 
Suppose there is a family $P$ of probability distributions defined on $\{0,1\}^n$ and there is a randomised mapping
  $\phi : \{0,1\}^n \rightarrow \{0,1\}^{n^*}, n^*$ $ \ge n$ for which there exists a mapping $\phi^{ -1}$
and let 
\begin{equation}\label{th2}\sup_{p \in P} (n^* - h_{min}(\pi_p))  \le \Delta \, . \end{equation} for some $\Delta$.
Then  

i)  $cipher_{c\&r}$ is $\epsilon$-entropically secure with secret key length $\Delta +2\log (1/\epsilon) +2$, and 

ii) $cipher_{c\&r}$ is $\epsilon$-indistinguishable with secret key length $\Delta +2\log (1/\epsilon) +6$.
}

Now we consider  a simple  example to illustrate the basic idea. Let $n=2$, $p(00) = 1/2, p(01) = 1/4$, $p(10) = p(11) = 1/8$. Obviously, $h_{min}(p) = 1$ and  $\Delta = (2 -1) $. The  map $\phi$  is constructed in two steps: first, "compress" the letters till $- \log p(a)$, that is, in our example, $00  \rightarrow 0$, $01   \rightarrow 10$ and $10   \rightarrow 110,  11  \rightarrow 111$. Secondly, randomise as follows: $00 $ uniformly $ \rightarrow \{000, 001, 010, 011 \}$,  $10  \rightarrow \{ 100, 101\}$ and two last letters as   $\{ 110\} $ and  $  \{111\}$ correspondingly.
As a result, we obtain a set $\{0,1\}^3$ subject to a uniform distribution whose min-entropy is equal to  three, and hence   $\Delta = 3 -3 = 0$.   
Thus, the key length becomes 1 bit shorter, but the message length is longer. It is proved that such a "bloated" cipher is $\epsilon$-entropically secure  \cite{rya-new}.

Obviously, the key length depends on the efficiency of the compression method, or code. Thus, in the case of known statistics (i.e., known $p$), the key length is $\Delta + 2  log (1/\epsilon) +2$, where $\Delta $ is 1 or 2 and depends  on the compression code chosen. If $p$ is unknown, but the messages are known to be generated by a Markov chain with known memory, then 
$\Delta = O(\log n)$ (and 
the key length is $O(\log n) + 2  log (1/\epsilon) $ \cite{rya-new} ).

\subsection{Universal coding}

The problem of constructing a single code for multiple probability distributions (information sources) is well known in information theory, and there are currently dozens of effective universal codes based on different ideas and approaches.  
It is worth noting that, at present, there are dozens universal codes, which are the basis for so-called archivers (e.g., ZIP). The first universal code for Bernoulli and Markov processes was proposed by Fitinghof \cite{ft}, and then Krichevsky found an asymptotically optimal code for these processes \cite{kr1,kr}. 
Other universal codes include the PPM universal code \cite{grev}, which is used together with the arithmetic code \cite{ri}, the Lempel-Ziv (LZ) codes \cite{ziv}, the Burrows-Wheeler  transformation \cite{bur}, which is used together with the book-stack code (or MTF) \cite{bs} (see also also \cite{mtf,acm}), grammar codes \cite{gb,gb2} and some others \cite{re,sp,tu,re2}. 

The universal code $c$ has to``compress'' sequences $x = x_1 ...  x_n$ that obey the distribution $p \in {\bf P}$ down to
 Shannon entropy $p$, that is $h_{Sh}(p)$, and the difference between $E_p(|c(x)|) - h_{Sh}(p)$ is called  redundancy $r(p)$  \cite{kr}  (here $E_p$ is the expectation and $|u|$ is the legth $u$). 
In \cite{u1}, an algorithm was proposed to construct a code $c_{opt}$ whose redundancy is minimal on ${\bf P}$, that is, $r_{p_{opt}} = \inf_{p \in {\bf P} } r(p) $. 
In \cite{u1} it was shown that $r_{p_{opt}}$ is equal to the capacity of a channel whose input alphabet is ${\bf P}$, whose output alphabet is the alphabet on which distributions from ${\bf P}$ are defined (in our case it is the alphabet $\{0,1\}^n$), and the lines of the channel matrix are probability distributions from ${\bf P}$ (see also \cite{u2} for the history of this discovery). This fact is important, because it allows us to use known  methods to compute the channel capacity to find the optimal code. 

In this paper, we will use the so-called Shtarkov maximum likelihood code $c_{Sht}$  \cite{sht}, whose construction is much simpler, and its redundancy is often close to that of the optimal code. This code is described as follows: first define
\begin{equation}\label{dfq}
p_{max}(u) = \sup_{p \in {\bf P} } p(u), u \in \{ 0,1 \}^n , \,\,
S_{\bf P} = \sum_{u \in \{ 0,1 \}^n } p_{max}(u),  \,\,q(u) = p_{max}(u)/S_{\bf P} .
\end{equation}
Clearly, 
\begin{equation}\label{q}
\forall  u \,: \, 
p(u)/ q(u) \le S_{\bf P} .
 \end{equation}

Shtakov proposed to build code $c_{Sht}$ for which 
$ |c_{Sht}(u)| =   \lceil  - \log q(u)  \rceil  . $
(Such a  code  exists,  see \cite{co}. )

Note that for a finite set $P$ \\ $S_P \le |P| $ (In particular, this is true when $P$ contains probability distributions corresponding to several languages).

\section{The cipher}

Now  we are going to construct an $\epsilon$-entropically secure cipher $c_{c\&r}$ for the case of unknown statistics, i.e., there exists some set of probability distributions {\bf P} generating words from $\{0,1\}^n, n \ge 1,$ and the constructed cipher should be applicable to messages obeying any $p \in {\bf P}$ with leakage no larger than $\epsilon$.
In short, we apply the general method from \cite{rya-new} to the probability distribution $q$ (\ref{dfq}). In detail,   
Alice wants to send messages $m \in  \{0,1\}^n$ to Bob, and they both know in advance that $m$ can obey any probability distribution $p$ of the set of distributions $ {\bf P}$.  The cipher algorithm is as follows.

{\bf Constructing the cipher.}\label{ps}
 We describe all calculations  in the following steps:

i)  compute the distribution $q$ according to (\ref{dfq}) and order the set $q(u), u \in \{0,1\}^n$. (Denote the ordered probabilities as $q_1, q_2,..., q_N$, $N = 2^n$   and let $\nu(u) = i$ for which $q(u) = q_i$.) 

ii) encode the ``letters'' $1, 2, ... , N$  with the distribution $q$ by the trimmed Shannon code from \cite{rya-new} .  Denote this code $\lambda$  and note that  
\begin{equation}\label{la}
\forall   i \,: \, 
| \lambda (i) | < -  \log q_i +2 \, 
 \end{equation}
and $\lambda$ is prefix-free, that is, for any $i$ and $j$, $i \ne j$,  neither $\lambda(i)$ is a prefix   $\lambda(j)$, no $\lambda(j)$ is a prefix $\lambda(i)$   \cite{rya-new}.

iii)  build the following  randomised map $\phi$ 
 First, find $n^* = \max_i \lambda(i) $ and then define for $u \in \{0,1\}^n,  $
\begin{equation}\label{phi} \phi(u) = \lambda(\nu(u) ) r_{ |\lambda(\nu(u)| + 1}\,  ... \, r_{n^*} \, , 
 \end{equation} 
where $r_j$ are equiprobable independent binary digits.

iv) For the   desired   leakage $\epsilon$ build $cipher_{rw-ds}$ with secret key length 
\begin{equation}\label{klength}
\lceil \log S_{\bf P}  \rceil+ 2\log (1/\epsilon) +\delta \, ,
\end{equation}
where $\delta = 2$ for $\epsilon$-entropically secure cipher and $\delta =6$ for $\epsilon$- indistinguishable one.

It is worth noting that Alice and Bob (and Eve) can do all the calculations described independently of each other.

{\bf Use of the cipher. }
Suppose Alice and Bob have a randomly chosen secret key $K$, $|K|= k $, and Alice wants to send Bob a message $m$. To do this, she computes $cipher_{c\&r}(m,K)$, as described above, and sends it to Bob.

Bob receives the word $cipher_{c\&r}(m,K)$ and decrypts it with the key $K$. As a result he gets the word $ \phi(m) = \lambda(\nu(m ) r_{ |\lambda(\nu(m)| + 1}\,  ... \, r_{n^*} \,$ whose prefix $\lambda(\nu(m )) $ defines the message $m$ (this is possible because $\lambda$ is prefix-free).  

The properties of this cipher are described in the following theorem.

{\bf Theorem 3.}
Suppose there is a family $P$ of probability distributions defined on $\{0,1\}^n$ and 
some $\epsilon > 0$.  If the described $cipher_{c\&r}$ is applied then

i) the  $cipher_{c\&r}$ is $\epsilon$-entropically secure with secret key length $\lceil \log S_{\bf P}  \rceil+2\log (1/\epsilon) +2$ and 

ii) the $cipher_{c\&r}$ is $\epsilon$--indistinguishable with secret key length $\lceil \log S_{\bf P}  \rceil +2\log (1/\epsilon) +6$.

{\it Proof.} 
For any $p \in  {\bf P}$  the random map $\phi$ defines a probability distribution $\pi_p(v), v \in \{0,1\}^*$ as follows:
for    any $u \in \{0,1\}^n $ and         $v \in \phi(u)  $   
$$
\pi_p(v) = p(u) 2^{-(n^* -| \lambda(\nu(u)|)} \, ,
$$
see (\ref{phi}).
From definitions $\phi$ and (\ref{phi}),   (\ref{la}) we obtain 
$$ \pi_p(v) = p(m) 2^{ -({n^*   -   |\lambda(\nu(m)|)}}  \le  p(m) 2^{-( n^* -(\log q_{\nu(m)} +2))}  $$ 
for any $m \in   \{0,1\}^n $ and $v \in    \phi(m) \subset  \{0,1\}^{n^*}$.   Then 
$$
- \log  \pi_p(v) \ge - \log p(m)   -(n^* - (\log q_{\nu(m)} +2))  \ge $$ $$ \log S_{\bf P} - \log q_{\nu(m)}- (n^* - (\log q_{\nu(m)} +2)) = \log S_{\bf P} + 2  - n^* \, 
$$ for any $m$ and $v \in  \phi(m) \subset  \{0,1\}^{n^*}$.
So,  $h_{min} (\pi_p)  =  \min_{v \in \{0,1\}^{n^*}}  - \log  \pi_p(v ) \ge \log S_{\bf P} + 2  - n^* $ and, hence, 
  $\sup_{p \in {\bf P} } (n^* - h_{min} (\pi_p) ) \le \log S_{\bf P} + 2 $ .  From   (\ref{th2}) (Theorem 2) and the description of the cipher (\ref{klength})    we can see that the $cipher_{c\&r}$  is

i)  $\epsilon$-entropically secure
with a secret key of length $\lceil \log S_{\bf P}  \rceil+ 2 \log (1/\epsilon) +4$ and

ii)   $\epsilon$-indistinguishable 
with a secret key of length $\lceil \log S_{\bf P}  \rceil+ 2 \log (1/\epsilon) +8$.

\section{ Conclusion}
We described the cipher for a family of probability distributions ${\bf P}$ defined on the set $\{0,1\}^n, n\ge 1, $ for which the length of the secret key does not depend directly on $n$, but depends on  ${\bf P}$.
 For example, if ${\bf P}$ is finite, the key length is less than $\log |{\bf P}| + 2 \log (1/ \epsilon) +O(1)$ and hence independent of $n$. This example includes the case where one needs to have the same cipher for texts written in different languages.
  Here, the size of the set ${\bf P}$ is equal to the number of languages. Thus, in some practically interesting cases, the extra length of the secret key is quite small.


\end{document}